\documentclass[%
 aip,
 amsmath,amssymb,
 reprint,%
]{revtex4-1}

\usepackage{graphicx}
\usepackage{dcolumn}
\usepackage{bm}

\usepackage[utf8]{inputenc}
\usepackage[T1]{fontenc}
\usepackage{mathptmx}

\begin{document}

\preprint{AIP/123-QED}

\title[]{Magnetization around mix jets entering inertial confinement fusion fuel}

\author{James D. Sadler}
 \email{james4sadler@lanl.gov}
 \affiliation{%
Los Alamos National Laboratory, P.O. Box 1663, Los Alamos, NM 87545, USA}%

\author {Hui Li}
\affiliation{%
Los Alamos National Laboratory, P.O. Box 1663, Los Alamos, NM 87545, USA}%
\author {Brian M. Haines}
\affiliation{%
Los Alamos National Laboratory, P.O. Box 1663, Los Alamos, NM 87545, USA}%

\date{\today}

\begin{abstract}
Engineering features are known to cause jets of ablator material to enter the fuel hot-spot in inertial confinement fusion implosions. The Biermann battery mechanism wraps them in self-generated magnetic field. We show that higher-Z jets have an additional thermoelectric magnetic source term that is not present for hydrogen jets, verified here through a kinetic simulation. It has similar magnitude to the Biermann term.  We then include this in an extended magneto-hydrodynamics approach to post-process an xRAGE radiation-hydrodynamic implosion simulation. The simulation includes an accurate model for the capsule fill tube, producing a dense carbon jet that becomes wrapped in a 4000\,T magnetic field. A simple spherical carbon mix model shows that this insulates the electron heat conduction enough to cause contraction of the jet to an optically thick equilibrium. The denser magnetized jet hydrodynamics could change its core penetration and therefore the final mix mass, which is known to be well correlated with fusion yield degradation. Fully exploring this will require self-consistent magneto-hydrodynamic simulations. Experimental signatures of this self-magnetization may emerge in the high energy neutron spectrum.
\end{abstract}

\maketitle

%
\section{Introduction}
                       
Inertial confinement fusion implosions are designed to create a $30$ micron sized plasma that is sufficiently hot and dense to cause rapid nuclear fusion of the deuterium and tritium fuel. If the temperature, density and size of the hot-spot exceed the criteria for ignition, then the fusion energy deposition will dominate the energy balance and cause a burn wave to propagate out into the cold surrounding fuel. This should lead to burn of a significant fraction of the several milligrams of fuel, leading to a MJ scale yield in national ignition facility experiments \cite{hurricane2014fuel}. However, the main factor limiting the achievable hot-spot conditions is unintended entry of contaminants into the fuel from the capsule fuel filling tube\cite{pak2020impact, ma2017role, PhysRevLett.111.045001, hammel2011diagnosing}. This tends to form a jet of heavier elements into the fusion core \cite{weber2020mixing}, composed of the fill tube material and surrounding ablator. Since the hot-spot is mostly optically thin to bremsstrahlung, x-ray radiation is a major loss source from the core plasma. Electron-ion free-free emission rates increase proportional to $Z^3$, meaning that heavier elements will increase the radiative losses and quench the burning plasma \cite{haines2020observation}. Furthermore, the cooling of the mix region will cause it to contract and increase in density, changing its surface area, total heat conduction and optical depth properties \cite{weber2020mixing}. 

Entry of carbon ablator spikes into the hot-spot was also predicted with three-dimensional radiation-hydrodynamic simulations of perturbed national ignition facility implosions \cite{clark2016three}. This occurs due to non-linear fluid instabilities. Depending on laser and target conditions, up to $200\,$ng of carbon can enter the fuel \cite{weber2020mixing}, constituting around 5\% of the hot-spot mass. This was studied experimentally through spectroscopy and by imaging localized regions of increased x-ray self-emission \cite{pak2020impact,  pickworth2018development, smalyuk2019review, bachmann2020localized}. Although there is significant experimental uncertainty on the contaminant mass, the fraction of x-ray yield from contaminants is better constrained. Fusion yield degradation was found to be linearly correlated with this fraction \cite{pak2020impact}, with the fraction reaching 50\% in some cases.

Although the loss of fusion yield due to increased radiative cooling has been confirmed experimentally, there may be other consequences of implosion asymmetries. Extended magneto-hydrodynamic (ExMHD) modelling of perturbed fusion implosions found that significant magnetic fields can be generated through the collisionless Biermann mechanism \cite{walsh2017self, walsh2019perturbation}. This has not been considered within the radiation-hydrodynamic modelling framework and is difficult to diagnose experimentally. 

In this work, we show that magnetization effects are greater for the fill tube jet than for pure hydrogen asymmetries, since the jet penetrates into hotter plasma. There is also an additional thermoelectric magnetic source term in multi-species plasmas. This is due to the $Z$ dependence of the Coulomb collision operator. We derive this thermoelectric magnetic source term and verify it via an electron Vlasov-Fokker-Planck kinetic simulation. Post-analysis of a perturbed radiation-hydrodynamic simulation shows that this source term has similar magnitude to the Biermann fields, which are themselves enhanced due to the increased temperature gradients from radiative cooling in the carbon jet. 

Magnetic fields are important in the hot-spot since they can affect the hydrodynamics through alterations to the electron heat flux. This is true even though the magnetic energy density is over $1000$ times less than the plasma energy density, such that ideal MHD would predict minimal coupling of the magnetic field back to the fluid. These effects are especially important in the time leading up to stagnation, when the plasma is sufficiently hot to be magnetized but not yet hot enough for fusion, so that electron heat flux dominates over alpha ion heat flux. As such, the ExMHD model is required, including the effects of anisotropic heat flux and non-ideal field advection. In this work, the magnetized reduction of electron heat conductivity into a micron size carbon jet reaches up to $70\,\%$ in the hundreds of picoseconds before stagnation. This is compared to around $20\,\%$ at the hot-spot edge perturbations. 

Due to the low magnetic pressure, the fields are not expected to couple back to the hydrodynamics in these implosions until close to fuel stagnation. This means a post-processing approach of a radiation-hydrodynamic simulation is sufficient to estimate the magnitude and profile of the magnetic fields near stagnation time.  Since we observe that the magnetic fields become significant, we then evaluate the coupling of the magnetic field back to the thermal transport. However, simulating the self-consistent evolution of the magnetized mix jet is left to future work. The reduced conduction is likely to increase the temperature gradients, meaning the true magnetic reduction of heat flux is less than the post-processed results presented here.

As the field strength $\mathbf{B}$ increases, electron heat flux is reduced transverse to the magnetic field and gains a component perpendicular to both the field and temperature gradient, known as the Righi-Leduc flux. These effects become significant when the magnetization Hall parameter $\chi=\omega\tau$ becomes similar to one, where $\omega=e|\mathbf{B}|/m_e$ is the electron gyro-frequency and 
\begin{align}
\tau &= \sqrt{\frac{9\pi}{2}}\frac{4\pi\epsilon_0^2\sqrt{m_e}(eT_e)^{3/2}}{n_e\tilde Z e^4\ln(\Lambda)}\label{tau}
\end{align}
is the electron Coulomb collision time. In this expression, $\epsilon_0$ is the vacuum permittivity,  $m_e$ is the electron mass, $T_e$ is the electron temperature in electron-volts, $n_e$ is the free electron number density, $e$ is the elementary charge and $\ln(\Lambda)$ is the Coulomb logarithm. The average ion charge state is defined through a sum over ion species of $\tilde Z(t, \mathbf{x})=\sum_jn_jZ_j^2/n_e$, where $Z_j$ is the species charge state and $n_j$ is the species number density. The parameter $\chi$ is also equivalent to the ratio of the electron Coulomb mean free path with its gyro-radius. With typical inertial confinement fusion hot-spot conditions, $\tau\simeq 1\,$fs, requiring fields of more than $5000\,$T to reach magnetization $\chi=1$. Due to the temperature dependence, magnetization corrections are most likely in the hotter plasma regions, precisely where fusion is occurring.

Due to the increased radiative cooling of the localized mix regions, there is a large inflow of electron heat flux. Magnetic reduction of this heat flux will reduce the equilibrium electron temperature in the mix region, especially in the time before alpha particle energy transport becomes dominant. This will increase the density of the spike and directly impact its hydrodynamics and radiative loss. 

In the next section of this work, we explore the additional collisional thermoelectric source of self-generated magnetic field, and then in section three we verify it with a kinetic simulation. In section four, we use the ExMHD model to post-process a radiation hydrodynamic simulation. In section five, we examine an energy balance model to estimate the effects of the magnetized heat flux on the mix jet dynamics. 

\section{Additional magnetic source term}

The additional magnetic field source term can be derived from the generalized Ohm's law. The detailed ExMHD electric field is given by \cite{epperlein1986plasma, molvig2014classical}
\begin{align}
    \mathbf{E} &= -\mathbf{u}\times \mathbf{B} + \frac{\mathbf{J\times B}}{n_ee} - \frac{\nabla p_e}{n_ee} + \underline\eta.\mathbf{J} - \underline\beta.\nabla T_e.\label{ohm}
\end{align}
This form has been verified by detailed kinetic simulations \cite{epperlein1986plasma, yin2016plasma, kingham2004implicit}. It assumes a quasi-neutral plasma with minimal kinetic transport, requiring that the Knudsen number $K_n=v_{th}\tau|\nabla T_e|/T_e\ll 1$, where $v_{th}=\sqrt{2eT_e/m_e}$ is the electron thermal velocity. This is satisfied in inertial confinement fusion hot-spots, where typically $K_n\simeq 0.005$, meaning there are only small kinetic corrections to the MHD description and Biermann term \cite{sherlock2020suppression}. Previous work also showed that there is a kinetic reduction to the fusion reactivity\cite{albright2013revised, sadler2019kinetic}, exceeding $10\,\%$ in regions with steep gradients. 

The first two terms in eq. (\ref{ohm}) are due to the relativistic transformation of the electric field from the electron fluid rest frame, calculated from the fluid velocity $\mathbf{u}$ and the current density $\mathbf{J}$.  The third term balances the gradient in electron pressure $p_e$ and leads to the usual Debye shielded electric field in a quasi-neutral plasma. For sub-sonic flows, the effects of electron inertia can be neglected. 

On ion hydrodynamic timescales, we make the standard magneto-hydrodynamics approximation to retain only low frequency oscillations, meaning that the displacement current is negligible in the Maxwell equations, giving $\mathbf{J}=\nabla\times\mathbf{B}/\mu_0$. This approximation eliminates the propagation of light waves and electron plasma waves. 

The final two terms in eq. (\ref{ohm}) are due to the collision operator. The plasma resistivity tensor $\underline\eta$ describes the electric field required to drive plasma currents in the presence of Coulomb collisions. The electron Coulomb-logarithm in the hot-spot is $\ln(\Lambda)=\ln(4\pi n_e\lambda_D^3/\tilde Z)\simeq 5$, sufficiently large that the collision operator is dominated by small angle Coulomb collisions, and therefore $\underline\eta$ can be modelled with the classical Fokker-Planck transport coefficients for weakly coupled plasma \cite{epperlein1986plasma}. 

The final term in equation (\ref{ohm}) is the thermoelectric or collisional thermal force. The thermoelectric term arises from the decreased collisionality of faster electrons from the hotter region of a temperature gradient. Its coefficient is therefore dependent on the electron-ion Coulomb collisionality, which increases with $\tilde Z$ [eq. (\ref{tau})]. 

The coefficient tensor $\underline{\beta}$ is dimensionless. In the case of zero magnetic field, it reduces to a simple scalar $\beta_0(\tilde Z)$. The magnitude of $\beta_0$ can be estimated from a simple physical argument. Taking an electron with typical thermal velocity $v_{th}$, its electron-ion collisional acceleration is $a\simeq v_{th}/\tau\propto T_e^{-1}$. This acceleration will vary for electrons arriving from hotter or colder plasma, leading to a net collisional acceleration $\delta a = -a\,\delta T_e/T_e=-\delta T_ev_{th}/(\tau T_e)$. The electrons arriving at the position of the ion can be assumed to originate from one collisional mean free path away. A spatial Taylor expansion of $T_e$, taking the distance as equal to the Coulomb mean free path $v_{th}\tau$, gives $\delta T_e \simeq v_{th}\tau|\nabla T_e|$. This leads to
\begin{align}
\delta a\simeq -\frac{v_{th}}{\tau T_e}v_{th}\tau|\nabla T_e| = -\frac{2e}{m_e}|\nabla T_e|. 
\end{align}
This collisional force must be balanced by an appropriate electric field $\mathbf{E}=(m_e/e)\delta a = -2\nabla T_e$. This demonstrates that the coefficient $\beta_0$ is of order one. It was calculated in eq. (98) of reference \cite{molvig2014classical} to be
\begin{align}
\beta_0(\tilde Z)\simeq \frac{30\tilde Z (15\sqrt{2}+ 11\tilde Z)}{288+604\sqrt{2}\tilde Z + 217\tilde Z^2},
\end{align}
increasing monotonically from $0.7$ to $1.5$. This form is in agreement with the numerical fit to the Fokker-Planck simulations in reference \cite{epperlein1986plasma}.

On substitution of equation (\ref{ohm}) into the Maxwell equation $\partial_t\mathbf{B}=-\nabla\times\mathbf{E}$, using $\nabla.\mathbf{B}=0$ and $p_e=n_eeT_e$, this results in the ExMHD induction equation\cite{walsh2020extended, sadler2020conference}
\begin{align}
\begin{split}
  \frac{\partial\mathbf{B}}{\partial t} =\,&\nabla\times(\mathbf{u_B}\times \mathbf{B}) + \eta_0\nabla^2\mathbf{B} - \nabla\eta_0\times(\nabla\times\mathbf{B})\\
  &-\frac{\nabla n_e\times\nabla T_e}{n_e} + \nabla\beta_0(\tilde Z)\times\nabla T_e.\label{induction}
\end{split}
\end{align}
The magnetic field is advected by the first term, with advection velocity $\mathbf{u_B}$ given by
\begin{align}
\begin{split}
  \mathbf{u_B} =\, &\mathbf{u} - (1+\delta_\perp)\frac{\mathbf{J}}{n_ee} + \delta_\wedge\frac{\mathbf{J\times\hat b}}{n_ee} \\&-\frac{e\tau}{m_e}\left(\gamma_\perp\nabla T_e + \gamma_\wedge\mathbf{\hat b\times\nabla }T_e  \right).
  \label{ub}
  \end{split}
\end{align}
It is composed of the ideal advection with the fluid velocity $\mathbf{u}$, plus some ExMHD corrections \cite{walsh2020extended} given in terms of the magnetic field direction $\mathbf{\hat b} = \mathbf{B/|B|}$. The $\delta(\tilde Z, \chi)$ and $\gamma(\tilde Z, \chi)$ transport coefficients are defined and plotted in reference \cite{walsh2020extended}. The $\delta_\perp$ and $\delta_\wedge$ terms give a small correction to the Hall velocity $\mathbf{-J}/(n_ee)$. The Nernst terms, on the second line of eq. (\ref{ub}), advect the field down temperature gradients, along with the electron heat flow. The cross-gradient Nernst term advects the field along isotherms, in the direction of $\nabla T_e\times\mathbf{\hat b}$. The Hall terms are only of the order $100\,$ms$^{-1}$ for the conditions discussed in this work, compared to $10^{5}$ms$^{-1}$ for the fluid and Nernst velocities. 

The first term in equation (\ref{induction}) gives the advection of the field, and the second its resistive diffusion and dissipation, with diffusivity $\eta_0=m_ec^2\epsilon_0\alpha_0/(n_ee^2\tau)$. The unmagnetized resistive transport coefficient $\alpha_0(\tilde Z)$ is given in reference \cite{epperlein1986plasma}. In high magnetic Reynolds number flows, such as inertial fusion hot-spots with $R_M\simeq 10^3$, the advection dominates over the resistive terms. 

On the second line of eq. (\ref{induction}), there are two magnetic field source terms that are still active when $\mathbf{B}=0$. The Biermann term operates on misaligned density and temperature gradients. There is also the thermoelectric source of magnetic field, given by misaligned gradients in ion charge state and electron temperature. Although $\beta_0$ is maximal for high $\tilde Z$ plasma, its derivative is maximal for $\tilde Z=1$, giving $\beta_0'(\tilde Z=1) \simeq 0.3$. Therefore for a low Z plasma with steep ion composition gradients, the thermoelectric term is of similar magnitude to the Biermann term and will be important in inertial confinement fusion conditions. Self-generated magnetic fields resulting from composition gradients were previously mentioned by M. G. Haines\cite{haines1986, haines1997}, but not fully explored in large-scale simulations of inertial confinement fusion. 

For the purposes of numerical simulation, it is often assumed that the Hall and Nernst velocities are negligible, the resistivity is spatially uniform and there is only a single ion species, such that $\nabla\tilde Z=0$. This results in the simplified resistive MHD model
\begin{align}
  \frac{\partial\mathbf{B}}{\partial t} &= \nabla\times(\mathbf{u}\times \mathbf{B})  + \eta_0\nabla^2\mathbf{B} - \frac{\nabla n_e\times\nabla T_e}{n_e}.
    \end{align}
However, this simplification is not appropriate in inertial confinement fusion plasmas, since there are large temperatures, steep gradients, Nernst velocities exceeding the fluid velocity and there are several different ion species. This requires the use of the full ExMHD model.

In terms of the dimensionless $\kappa(\tilde Z, \chi)$ heat transport coefficients, the magnetized electron heat conduction $\mathbf{q}$ is given by \cite{epperlein1986plasma}
\begin{equation}
    \mathbf{q} = -\frac{n_ee^2T_e\tau}{m_e}\left(\kappa_0\mathbf{\hat b}(\nabla T_e.\mathbf{\hat b}) + \kappa_\perp\mathbf{\hat b\times(}\nabla T_e\mathbf{\times\hat b)} \right) + \mathbf{q}_{RL}\label{heatflux}
\end{equation}
\begin{equation}
    \mathbf{q}_{RL} = -\frac{n_ee^2T_e\tau}{m_e}\kappa_\wedge\mathbf{\hat b\times }\nabla T_e
    \label{kappa}
\end{equation}
This includes the unaffected $\kappa_0$ heat flux along the field lines and the reduced heat conduction $\kappa_\perp$ across the field lines. There is also the Righi-Leduc heat flux component $\mathbf{q}_{RL}$ along the isotherms, perpendicular to both the temperature gradient and magnetic field. Previous studies found that this deflects heat along the intruding spikes, towards the edge of the hot-spot\cite{walsh2017self}. Heat transport towards the hot-spot edge decreases the fusion yield and allows the spikes to penetrate deeper. In ICF conditions the heat flux due to the current $\mathbf{J}$ is a factor of $10^3$ smaller than the thermal conduction and it can be neglected. 

\section{Verification of the thermoelectric source term with a kinetic simulation}
To verify the additional collisional thermoelectric source of magnetic field, we used a fully kinetic Vlasov-Fokker-Planck-Maxwell approach, similar to that outlined in the OSHUN code \cite{tzoufras2011vlasov}. The ions were treated as a fixed, constant temperature background, with ion and electron temperature profile initialized as $T(x,y) = T_0(1 + 0.1\sin(2\pi y/L))\,$ with $T_0=2000\,$eV. The wavelength was equal to the square two-dimensional Cartesian spatial domain size $L=5\mu$m. To eliminate the Biermann battery term, the electron number density $n_e = 10^{25}\,$cm$^{-3}$ was uniform. This means that any magnetic field that arises is solely due to the thermoelectric $\beta_0$ term. The plasma contained a mixture of ion species. The deuterium species had $Z=1$ and number density profile $n_D = n_e(0.9+0.1\sin(2\pi x/L))$, and the  carbon species had $Z=6$ and $n_C= (n_e-n_D)/6$, giving quasi-neutrality and mass density $\rho\simeq 30\,$gcm$^{-3}$.  The electrons were treated with the kinetic description and the Fokker-Planck operator, using the electron current to directly integrate the Maxwell equations. The electron distribution function was expanded in spherical harmonics in velocity space \cite{tzoufras2011vlasov} and the expansion truncated at first order. This is valid for the test set-up, since the electron thermal velocity is far greater than the electron fluid velocity.

The spatial grid resolution was $0.28\,\mu$m, with periodic boundaries. The velocity space grid had resolution $4.4\times10^6\,$ms$^{-1}$. The collision operator used fourth order accurate numerical integrals for the electron distribution and the analytic moment integrals for electron-ion collisions, assuming Maxwellian ion distributions at fixed temperature. The Fokker-Planck collision operators for the isotropic and anisotropic parts of the electron distribution function were used as outlined in eqns. (38) and (39) of reference\cite{tzoufras2011vlasov}. These collision operators were calculated on a uniform velocity space grid using fourth order accurate finite differences. The whole system was integrated with an explicit fourth order Runge-Kutta method for $10\,$fs, greatly exceeding the $35\,$as electron plasma period and the $0.7\,$fs electron-electron collision time. The $5\,$as time-step was set by the explicit algorithm stability constraints and sufficiently resolved all of these time-scales. In addition, the $20\,$nm electron mean free path is far shorter than the gradient scale-lengths in the simulation, so eqn. (\ref{induction}) should be valid.

  \begin{figure}[t]
  \includegraphics{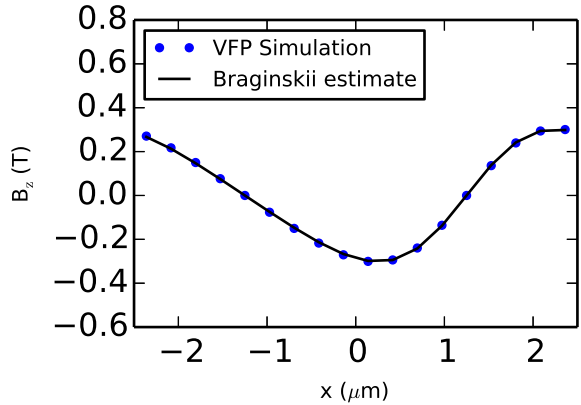}
  \caption{Verification of the additional thermoelectric magnetic source term with a two-dimensional Vlasov-Fokker-Planck-Maxwell code. The dots show the $y=0$ line-out of the $z$ component of the magnetic field after $10\,$fs, as calculated from the collisional kinetic simulation code. The solid line shows the ExMHD prediction of the final term in equation (\ref{induction}), multiplied by the $10\,$fs simulation duration. The ion charge state and electron temperature were initialized with a sinusoidal profile in the $x$ and $y$ directions respectively. The electron density was uniform, eliminating the Biermann battery term. }
   \label{vfp}
\end{figure}

A line-out of the $z$ component of the magnetic field at $y=0$ is shown in Fig. \ref{vfp} and compared with the theoretical estimate from the thermoelectric (final) term in eq. (\ref{induction}), multiplied by the simulation duration $10\,$fs. The kinetic magnetic field agrees well with the theoretical estimate of the thermoelectric term, which dominates over all other terms in eq. (\ref{induction}). Note that due to the non-trivial dependence of $\tilde Z$ and $\beta_0(\tilde Z)$, the magnetic field does not have a sinusoidal profile.

\section{ExMHD Analysis}

Numerical integration of the ExMHD induction equation (\ref{induction}) will give an indication of the magnetization induced as a result of carbon mix spikes entering the fusion hot-spot. We post-processed results from a two-dimensional xRAGE radiation-hydrodynamic simulation conducted in cylindrical coordinates. 

xRAGE is an Eulerian radiation-hydrodynamics code developed at Los Alamos National Laboratory \cite{gittings2008,haines2017}.  xRAGE features adaptive mesh refinement (AMR), enabling it to accurately model the fill tube geometry in high resolution.  Simulations include hydrodynamics, multi-group radiation diffusion for radiation transport, a three-temperature (electron, ion, and radiation) plasma model, electron and ion thermal conductivity, thermonuclear burn, SESAME tabular equations of state \cite{sesame}, and OPLIB opacities \cite{colgan2016}.  Simulations use the methodology for performing capsule-only simulations of indirectly driven capsule implosions outlined in reference\cite{haines2017}.  They are driven by a frequency-dependent boundary x-ray flux source derived from hohlraum simulations performed in HYDRA \cite{marinak2001}.

For the purposes of this study, we have performed simulations of NIF shot N170601\cite{lepape2018}, which was the first implosion at full NIF power fielded with a 5$\mu$m fill tube and achieved the first capsule yield above $10^{16}$ neutrons.  xRAGE simulations of this capsule have previously been reported in \cite{haines2020ft}.  Simulations were performed at a maximum AMR resolution of $0.25\mu$m and include accurate models for the fill tube, bore hole, and glue geometry assuming 2D axisymmetry based on as-shot characterization.  Simulations also include surface roughness based on capsule measurements.  xRAGE simulations were performed with interface preservers disabled, allowing numerical diffusion, so that they can be considered implicit large eddy simulations.  This strategy has resulted in favorable comparisons with simulations including explicitly modeled plasma transport models \cite{haines2020ft}.

The xRAGE code does not include magnetic fields. However, since the expected magnetic field pressure is far below the plasma pressure, the hydrodynamic results can be post-processed to estimate the resulting field values. If the magnetic field becomes large enough to produce $\chi\simeq 1$, it will start to affect electron heat conduction, at which point the magnetic field significantly alters the hydrodynamics and a full ExMHD simulation would be required\cite{walsh2017self}. 

  \begin{figure*}[t]
  \includegraphics{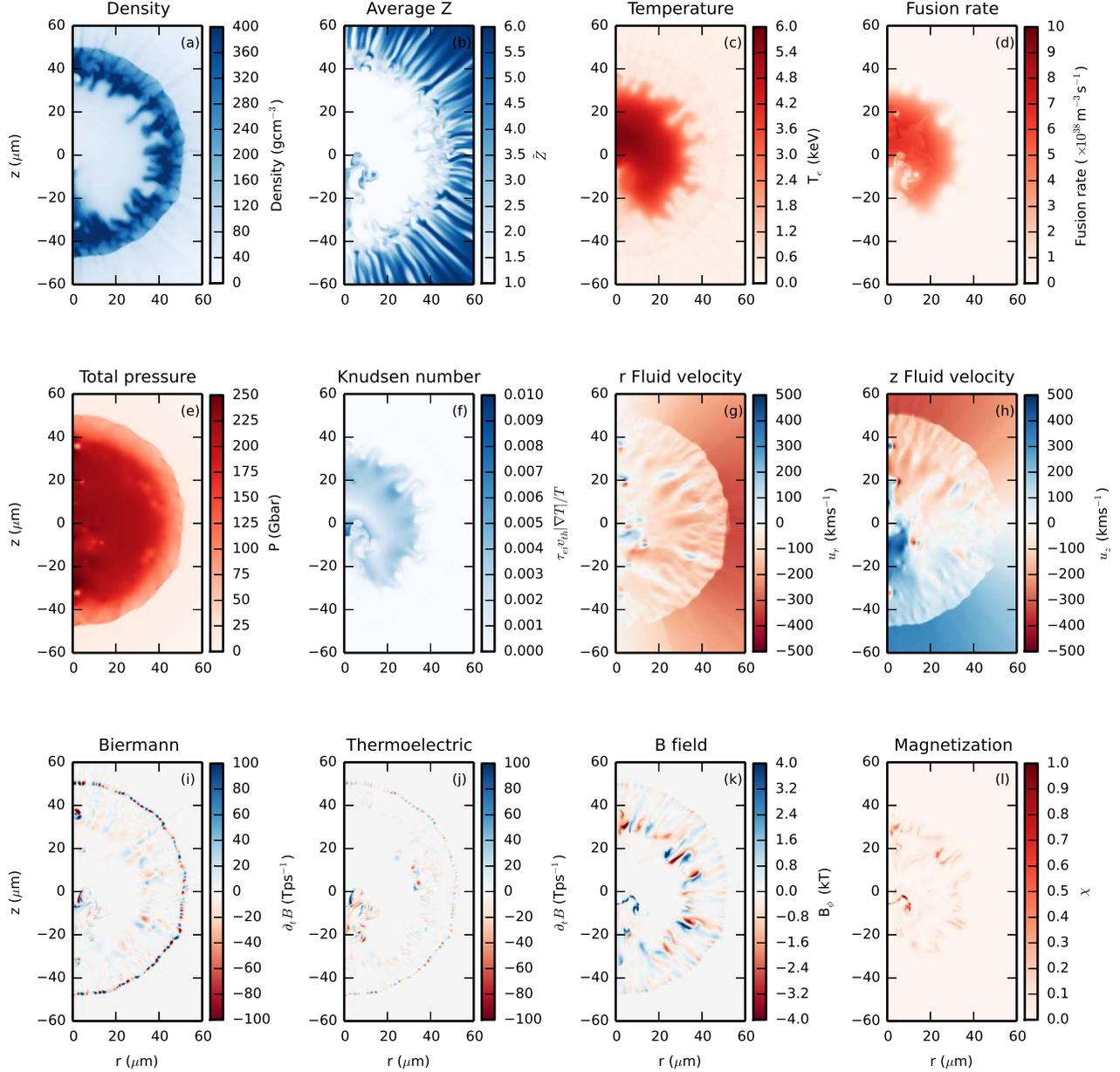}
  \caption{Results and ExMHD post-analysis of the two-dimensional cylindrical co-ordinates xRAGE radiation-hydrodynamic simulation of an inertial confinement fusion implosion, shown at near peak compression after $8300\,$ps. The fill tube jet is seen entering the hot-spot from negative $z$ and $r\simeq 0$. (a) The total mass density. (b) The effective ion charge state $\tilde Z$. (c) The electron temperature. (d) The volumetric fusion reaction rate. (e) The total pressure. (f) The Knudsen number $\tau v_{th}|\nabla T_e|/T_e$. Values above $0.01$ begin to show kinetic non-local changes to electron heat flux and ExMHD magnetic source terms. (g) The fluid velocity component in the radial $r$ direction. (h) The fluid velocity component in the $z$ direction. (i) The Biermann battery magnetic source rate. (j) The thermoelectric magnetic source rate. (k) The magnetic field, calculated by integrating eq. (\ref{induction}) using the fluid quantities over the preceding $300\,$ps. (l) The resulting electron magnetization.   }
   \label{xrage}
\end{figure*}

The magnetic field was integrated using equation (\ref{induction}). Due to the cylindrical coordinates geometry of the simulation and zero initial magnetic field, the electric field is directed in the plane of the simulation and so the self-generated magnetic field is in the azimuthal direction, out of the simulation plane. At the simulated fluid velocities, fluid elements move across more than 20 cells between each available xRAGE data-set, separated by $\Delta t=20\,$ps. Therefore a much shorter time-step of $\Delta t/100$ was used to numerically integrate equation (\ref{induction}) over each data time-step period from $t=t_0-\Delta t/2$ to $t=t_0+\Delta t/2$. The accuracy of each of these integration periods is severely restricted by continued use of the fluid quantities at $t=t_0$. Magnetic advection with velocity $\mathbf{u_B}$ used a conservative and positivity preserving flux corrected transport algorithm. The spatial grid resolution sufficiently resolved the fine features, such that second-order cylindrical-coordinate finite differences were sufficient to calculate the resistive and magnetic source terms in equation (\ref{induction}). Transport coefficients were calculated from the polynomial fits in Epperlein and Haines\cite{epperlein1986plasma}. Due to the magnetic advection with the imploding plasma, domain boundary effects were found to be negligible.

The results are shown in Fig. \ref{xrage}, which plots several simulation quantities at a time $20\,$ps before peak fusion burn. Panels (a-h) are direct output from the xRAGE simulation and panels (i-l) are from the magnetic field post-analysis. The implosion has created a hot-spot of density $50\,$gcm$^{-3}$ and peak temperature $5.9\,$keV. Perturbations have grown at the hot-spot to fuel shell and fuel shell to carbon ablator interfaces in the density plot Fig. \ref{xrage}a. This can also be seen in Fig. \ref{xrage}b, which shows the average ion charge state $\tilde Z$. The steep gradients in $\tilde Z$ are required for the thermoelectric source term outlined in eq. (\ref{induction}). The fill tube jet, composed of heavier elements, can be seen entering the hot-spot on-axis from negative $z$. It reaches the centre of the implosion. There are also steep electron temperature gradients, shown in Fig.\ref{xrage}c. The carbon mix jet region has a much lower temperature due to its increased radiative cooling, deforming the shape of the hot-spot and increasing its surface area. This will be explored further in the next section.

Fig. \ref{xrage}d shows the volumetric fusion reaction rate. There is very little fusion happening inside the mix jet region, partly because the fuel mass fraction is low here, and also due to the lower temperature. The mix jet region leads to increased x-ray emission \cite{pak2020impact}, so an offset between the regions of greatest x-ray and neutron production in the hot-spot images may indicate a substantial mix jet of the type simulated. Although the simulated x-ray drive was symmetric and the overall hot-spot is spherical, the mix jet region gives the fusion burn profile an oblate P2 asymmetry when convolved with a $10\,\mu$m imaging resolution. It has a shorter extent in the $z$ direction. Early-time shadowing from the fill tube will also contribute to this. It is important to note that a neutron image shape with oblate P2 asymmetry can therefore be caused by a single encroaching high-Z mix jet, as well as by drive asymmetries.  

An additional consideration is the kinetic reduction of fusion reactivity due to plasma gradients\cite{albright2013revised}, which can reduce the fusion yield even 5-10 microns away from the mix jet. This means the true reaction rate is likely to be $10-20\%$ less than this fluid calculation, especially in the regions around the mix jet at $z\simeq -20\,\mu$m.

As seen in Fig. \ref{xrage}e, the mix region remains in pressure equilibrium with the rest of the hot-spot. It cools and contracts, which also increases its optical depth. The validity of the magneto-hydrodynamics model and eq. (\ref{ohm}) is verified by the electron Knudsen number, shown in Fig. \ref{xrage}f. This is given by the ratio of the electron Coulomb mean free path with the temperature gradient scale-length. Values of $K_n<0.01$ indicate the validity of the magnetic source terms \cite{sherlock2020suppression}. However, the Knudsen number can reach $0.1$ in the DT gas at earlier times in the implosion. As such, the magnetic integration analysis only covered the preceding $300\,$ps period, in which eqn. (\ref{induction}) is valid.

Figs. \ref{xrage}g,h show the fluid velocity components. The carbon fill tube jet has a large speed of $400\,$kms$^{-1}$ inwards, reaching the centre of the hot-spot. This motion advects the magnetic field towards the hot-spot centre and leads to a large magnetic Reynolds number $R_M\simeq 10^3$. Although this means magnetic advection is dominant over diffusion, turbulent dynamo magnetic amplification is not expected due to the much lower fluid Reynolds number \cite{weber2020mixing, tzeferacos2018laboratory}. 

Fig. \ref{xrage}i shows the instantaneous Biermann magnetic field generation rate, and compares it to the thermoelectric generation rate in Fig. \ref{xrage}j. These have been calculated using a second order finite difference method. As expected, the two source terms are comparable in the fusion fuel, with large-scale regions at $50\,$Tps$^{-1}$ and several small regions producing $100\,$Tps$^{-1}$. There is significant field growth in the modulations at the edge of the hot-spot, as well as around the carbon jet that enters the fuel from negative $z$. The highest Biermann growth is also in a small vortex that arises opposite the incoming fill tube jet, located at $z=35\,\mu$m. 

The thermoelectric magnetic source term is predominantly around the fill tube jet, reaching peak values of $100\,$Tps$^{-1}$ within $100\,$ps of the stagnation time. The Biermann term is more prominent than the thermoelectric term at the DT ice to carbon ablator interface, and at the edge of the hot-spot.

The thermoelectric magnetic source is often in the opposite direction to the Biermann source. The jet radiatively cools and contracts, meaning $\nabla n_e$ is in the same direction as $\nabla \tilde Z$. However, the two terms have opposite sign in eq. (\ref{induction}) and so will be in opposite directions.

To assess the growth of the magnetic field, equation (\ref{induction}) was integrated only over the preceding $300\,$ps, starting with $\mathbf{B=0}$. As such, Fig. 
\ref{xrage}k is likely to show a conservative estimate of the self-generated magnetic field. Field generated at earlier times is compressed by its advection with the fluid, increasing the final field strength. However, there is also magnetic dissipation due to the resistivity. This stabilises the magnetic field magnitude and means the magnetic compression is less effective than the fluid compression. Despite this effect, the highest field values are around $4000\,$T at the edge of the carbon mix spike and across the larger perturbation region at the edge of the hot-spot. This value is in approximate agreement with the values calculated with the full ExMHD simulations in reference \cite{walsh2017self}. 

The Biermann source rate is also large at the DT ice to carbon ablator interface. However, the plasma is colder and denser here, leading to greater dissipation from the magnetic diffusivity $\eta_0$. As a result, magnetic field in Fig. \ref{xrage}k mostly accumulates at the edge of the hot-spot and not at the DT ice to ablator interface. The overall integrated $\mathbf{B}$ field around the mix jet continues to be in the direction of the Biermann source term, such that the Righi-Leduc heat flow $\propto \nabla T_e\times \mathbf{\hat b}$ remains towards the base of the encroaching spikes\cite{walsh2017self}. 

  \begin{figure}[t]
  \includegraphics{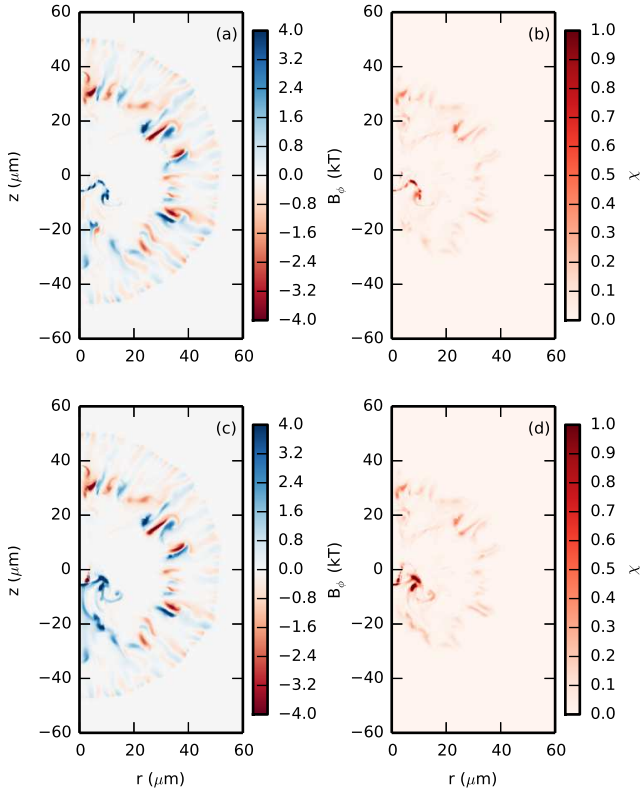}
  \caption{Integration of the magnetic field from the two-dimensional radiation-hydrodynamic simulation, shown at $t=8300\,$ps, for the cases of (a, b) Biermann and thermoelectric source terms, and (c, d) the Biermann source term only. Panels (a) and (c) show the integrated magnetic field using equation (\ref{induction}) and panels (b) and (d) show the resulting electron magnetization $\chi$. The thermoelectric term causes differences around the fill tube jet.}
   \label{comparison}
\end{figure}

Fig. \ref{xrage}l shows that the electron magnetization $\chi$ reaches values close to $1$ at the edge of the hot-spot and around the fill tube region. This is strong enough to reduce the heat flux magnitude by half and deflect its direction by $\simeq 45^\circ$ from the temperature gradient direction. The magnetization effects are predominantly around regions of high temperature gradients, precisely where electron heat flux is maximal.

To isolate the effect of the thermoelectric term, the analysis was repeated with it set to zero. The results at $t=8300\,$ps are shown in Fig. \ref{comparison}, where the top panels include both source terms and the bottom panels include the Biermann term only. The thermoelectric term causes the most differences around the carbon fill tube region. The resulting field with both source terms is still in the positive azimuthal $\phi$ direction, but reduced in extent and magnitude compared to the field accumulated from the Biermann term alone. Figs. \ref{comparison}b and \ref{comparison}d also indicate that the thermoelectric term reduces the magnetization around the fill tube jet, while leaving it relatively unchanged at the hot-spot edge.

  \begin{figure}[t]
  \includegraphics{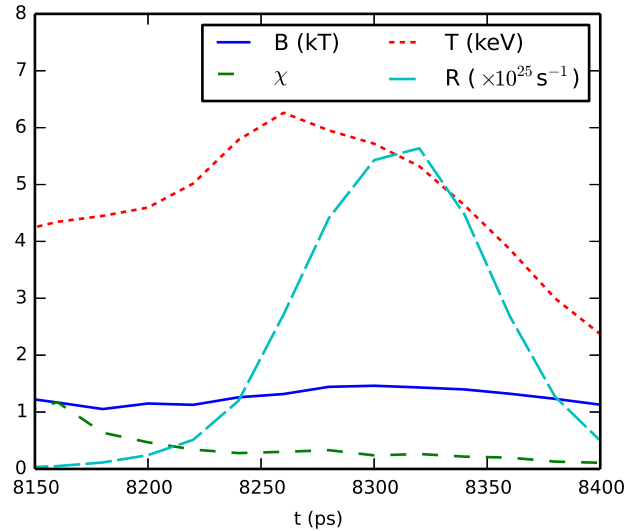}
  \caption{ Time evolution of quantities from the two-dimensional radiation-hydrodynamic simulation. The plot shows the root mean square magnetic field $|\mathbf{B}|$ in kiloTesla, the root mean square magnetization $\chi$, the maximum electron temperature value in keV and the volume integrated fusion reaction rate. The root mean square field is calculated over cells with $|\mathbf{B}|>500\,$T. The root mean square magnetization is calculated over cells with $\chi>0.1$. }
   \label{timetrace}
\end{figure}

  \begin{figure}[t]
  \includegraphics{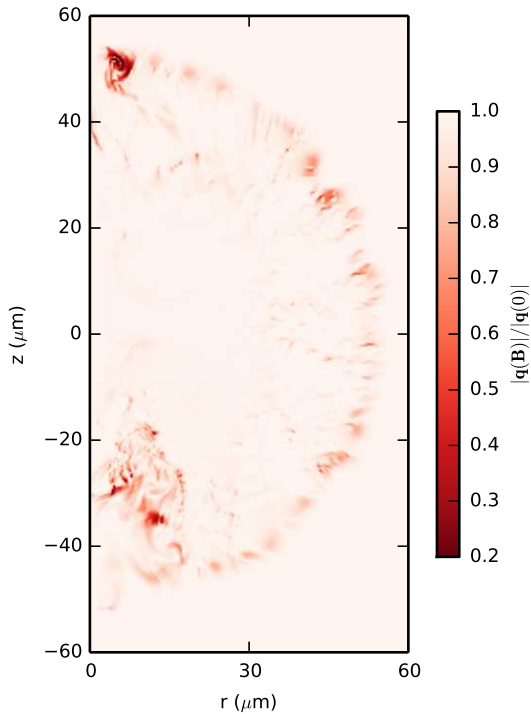}
  \caption{The magnetized electron heat flux magnitude, normalized to the case without $\mathbf{B}$ field. This is calculated from the ExMHD post-processing of the two-dimensional xRAGE radiation hydrodynamic simulation, shown at $8200\,$ps, around $120\,$ps before bang time. The self-magnetized fill tube mix jet is seen at $z\simeq -35\,\mu$m.    }
   \label{q}
\end{figure}

Fig. \ref{timetrace} shows the time evolution of the fusion burn and magnetic quantities. The fusion rate peaks at $t=8320\,$ps, with a burn width of $100\,$ps. The electron temperature reaches a maximum of $6.3\,$keV at a time slightly before this. The solid line shows the root-mean-square magnetic field, which is also maximal at around the time of peak burn. The field continues to grow over time due to the continuing magnetic source terms, magnetic compression, and Nernst advection compressing the field down the temperature gradients into the colder spike regions. However, the magnetization $\chi$ this produces actually decreases closer to bang time. This is due to the temperature and density dependence of $\chi$. The magnetization is reduced by the increased density at peak compression, and furthermore the field is Nernst advected into the cooler regions by the large heat fluxes around bang time. Compression of the field into the cooler, denser regions increases the maximal field strength, but decreases the magnetization since $\chi\propto T_e^{3/2}$.

Fig. \ref{q} shows the effect of the magnetization on the electron heat conduction. This is especially important for mix jet dynamics at earlier times, shown here at $t=8200\,$ps when the fusion burn has barely started and electron heat flux is dominant over alpha particle heat flux. At this earlier time, the carbon mix jet has already penetrated the hot DT core and so its magnetization exceeds one, higher than the value at the hot-spot edge. Magnetization $\chi>1$ reduces the heat conductivity by up to $70\,\%$ around the fill tube jet and the small vortex that arises on the opposite side.  This greatly exceeds the reduction around the edge of the hot-spot, which is nearer to $20\,\%$. For this reason, magnetization of the fill tube jet is more significant than the magnetic effects at the hot-spot edge, simply because it happens earlier. This will allow extra time for magnetized hydrodynamics in the jet.

If these magnetization effects are included self-consistently in the simulation, the reduced conductivity will reduce the electron heat flux, causing a steeping of the temperature gradient. The equilibrium magnetized heat flux can be estimated from the steady state heat diffusion equation $\nabla.(\kappa \nabla T_e)=f$, where $\kappa$ is the heat conductivity and $f$ is a spatially dependent source term. Dimensional analysis then leads to a temperature gradient scale-length $L=\sqrt{\kappa T_e/f}$ and therefore an estimated heat flux magnitude $|-\kappa\nabla T_e|=\kappa T_e/L = \sqrt{\kappa T_e f}$. This shows that any magnetic reduction in conductivity $\kappa$ will lead to a lesser reduction in the self-consistent heat flux, since $\mathbf{q}\propto \sqrt{\kappa}$ rather than the linearized version $\mathbf{q}\propto \kappa$ shown in Fig. \ref{q}. However, this steepening of the temperature gradient may also increase the Biermann and thermoelectric magnetic field source rates, increasing the magnetization above the post-processed values presented in this work. The combination of these competing effects will require full ExMHD simulations to find the correct self-consistent heat flux. The heat flux reduction presented in Fig. \ref{q} is therefore only an approximation to the true effect.

The parameters simulated in this section were chosen to be representative of the current high yield national ignition facility implosions. With consideration of eqns. (\ref{tau}) and (\ref{induction}), we see that the magnetic effects will be maximal for shots with higher core temperatures,  lower densities and steeper gradients. In addition, the thermoelectric source term will have increased prominence for jets of higher $Z$ material entering the hot core plasma. As such, these magnetization effects are expected to be most important for implosions with large surface defects, greater convergence ratio, higher temperature and lower hot-spot density. 

Although reference\cite{walsh2017self} found that the self-magnetization barely changes yield and hot-spot temperature, this conclusion did not consider its effects on the high-Z fill tube jet dynamics. Fig. \ref{q} shows that magnetization of the jet is much more significant in the time before stagnation. In the next section, we show that this has an indirect effect by altering the density and radiative losses of the fill tube jet. Changes to the jet density and hydrodynamics will also impact its eventual penetration into the hot-spot, altering the final hot-spot mix mass compared to the prediction from pure radiation-hydrodynamic simulations.

\section{Energy balance model}
Although the simulation did not include the effects of this magnetized heat flux, we may estimate its impact through generalising the mix jet energy balance outlined in reference\cite{weber2020mixing}. This simple model treats the mix region as a carbon sphere, with temperature $T_e$, embedded in a hot-spot at fixed temperature $T_0$ and pressure $P_0$. The energy of the mix region changes as a result of electron and alpha ion conduction at its boundary, radiative losses, and compression. This leads to the evolution of the energy of the mix region

\begin{align}
    \frac{\partial E}{\partial t} = Q_e + Q_\alpha - Q_{rad} - P_0\frac{\partial V}{\partial t},
\end{align}
where $V$ is the sphere volume. Assuming that the sphere remains isobaric with the hot-spot at $P_0$, as evidenced by Fig. \ref{xrage}e, $V$ can be found using the ideal gas law for the sphere $P_0V=NeT_e$. The energy balance becomes

\begin{align}
\frac{5}{3}\frac{\partial E}{\partial t} = \frac{5}{2}Ne\frac{\partial T_e}{\partial t} =  Q_e + Q_\alpha - Q_\mathrm{rad},\label{energymodel}
\end{align}
where $N$ is the total number of particles in the sphere. Electron and ion temperatures are assumed equal.
The electron heat conduction can be modelled using the $\kappa_\perp$ term from eqn. (\ref{heatflux}), where the cross field coefficient $\kappa_\perp\simeq\kappa_0/(1+2\chi^2)$ is the correct heat flux coefficient, since the magnetic field wraps around the mix region azimuthally. The conduction power $Q_e$ is this heat flux multiplied by the surface area of the sphere $4\pi r^2$. The sphere radius $r$ is found from $V$, whereas the temperature gradient scale-length is set to a fixed value of $r/2$. To approximate the boundary, the $n_e$, $T_e$, $\tau$ and $\chi$ values are evaluated using a geometric mean of the $n_e$ and $T_e$ inside and outside of the sphere. The interior electron density $n_e$ can be found using $N=N_C+N_e = 7N_e/6$ and $n_e=N_e/V$. In the fuel assembly period, at least $100\,$ps before bang time, the fusion rate is negligible and $Q_\alpha$ is negligible compared to $Q_e$.

At temperatures typical of fusion plasma, the carbon is fully ionised and the radiative term is dominated by the electron free-free bremsstrahlung. This has the form\cite{NRL}
\begin{align}
    Q_{rad} = c_b n_e^2\tilde Z\sqrt{T_e} V_{\sigma<1},
\end{align}
with $c_b=1.69\times 10^{-38}\,$eV$^{-1/2}$Wm$^{3}$. However, as noted in reference \cite{weber2020mixing}, this radiative loss is only from the volume from the sphere surface that has optical depth less than one, using the Bremsstrahlung optical depth in reference\cite{NRL}. As the mix region cools and contracts, $V_{\sigma<1}$ will decrease below $V$, limiting the radiative loss at low temperature. 

  \begin{figure}[t]
  \includegraphics{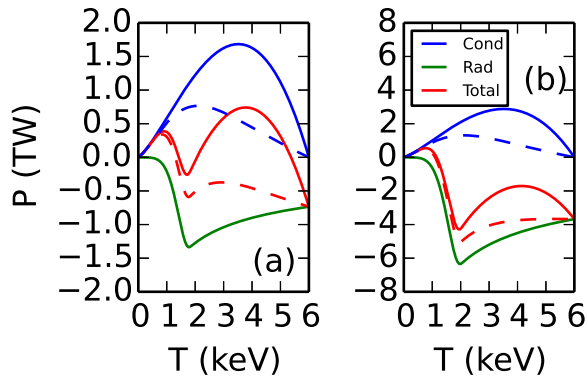}
  \caption{Rate of energy change for the mix region in the spherical carbon mix model, as a function of the carbon temperature. The plots show the conduction $Q_e$ (blue) and radiative loss $Q_{rad}$ (green) terms for a carbon sphere of mass (a) $20\,$ng and (b) $100\,$ng set inside a hydrogen hot-spot of fixed temperature $6\,$keV and pressure $175\,$Gbar, similar to the values from Fig. \ref{xrage}. The total of these two terms is shown in red. Solid lines are for $|\mathbf{B}|=0$ and dashed lines are for $|\mathbf{B}|=5000\,$T. }
   \label{fig:energy}
\end{figure}

Fig. \ref{fig:energy} shows eqn. (\ref{energymodel}) for $20\,$ng and $100\,$ng mix spheres set in a hot-spot of pressure $P_0=175\,$GBar and temperature $T_0=6\,$keV. When the mix has the same temperature as the hot-spot, electron conduction is zero but there is still radiative loss. At cooler temperatures $T_e\simeq 5\,$keV, the conduction into the sphere partially compensates for this in the $100\,$ng case. In the $20\,$ng case, it fully compensates for the radiative loss, such that the total $\frac{\partial T_e}{\partial t}=0$ at a stable equilibrium near $5\,$keV for the $\mathbf{B}=0$ case. In the $100\,$ng case, the conduction is not enough to balance out the radiative losses and the sphere keeps on cooling to below $4\,$keV, at which point the conduction gains start decreasing. This is partly because the lower temperature decreases the heat conductivity, and also because the mix region contracts and so its surface area decreases. The collisional Bremsstrahlung radiative rate initially increases as the mix region cools and contracts, since $n_e\propto 1/T_e$ and so $Q_{rad}\propto n_e^2\sqrt{T_e}\propto T_e^{-1.5}$. Below $T_e\simeq 2\,$keV, the radiative rate decreases due to the optical depth effect as the sphere becomes denser and cooler. This reduction in radiation causes the $100\,$ng sphere to reach equilibrium around $1.5\,$keV, in agreement with Fig. \ref{xrage}c.

The power rates shown in Fig. \ref{fig:energy} will cause local equilibriation of the mix region, within tens of picoseconds, to a temperature $T_e<T_0$ given by the zeros of the red curves. As shown by Fig. \ref{fig:energy}, mix regions exceeding a certain critical mix mass will quickly contract until they are optically thick, with $T_e\simeq 1.5\,$keV, whereas mix regions below this mass will tend to equilibriate at $T_e\simeq T_0$ where they are still optically thin and conduction balances the increased radiative loss from the carbon. This means that the energy loss from the hot-spot to the mix region is dominated by the steady-state conduction losses after equilibrium has been reached (hundreds of Joules), rather than the work done during contraction and the internal energy of the mix region (tens of Joules). Once equilibrium is reached at $Q_\mathrm{rad}=Q_e+Q_\alpha$, this means the power loss from the hot-spot to the mix region is simply $Q_{rad}$. From the zeros of the red curves in Fig. \ref{fig:energy}, the hot-spot loses energy to the mix region at a rate of $0.8\,$TW in the $20\,$ng case and $1.4\,$TW in the $100\,$ng case. The smaller mass sphere is therefore more damaging per unit mix mass\cite{weber2020mixing}.

The dashed lines in Fig. \ref{fig:energy} show the effect of an applied $5000\,$T magnetic field around the edge of the sphere. Again, this reduction is likely an over-estimate of the true effect, since the nonlinear feedback on the temperature profile is neglected. The field leads to values of $\chi$ as high as $1.8$, reducing the heat flux into the sphere by a factor of $1+2\chi^2$.  Magnetization $\chi$ increases with temperature, such that the fractional reduction in conduction is greatest for the highest temperatures in Fig. \ref{fig:energy}. The total energy gain of the sphere (red dashed curve) is now negative for the $20\,$ng case, right down to $T_e\simeq1.5\,$keV, so it will cool until it becomes optically thick at $T_e\simeq 1.5\,$keV, at which point the conduction balances its radiative loss. The magnetic field therefore has the effect of reducing the critical mix jet mass required for a collapse to an optically thick state. 

 This simple model identifies the existence of a critical mix jet mass, above which the mix region contracts to an optically thick state. For a given temperature, magnetization can only decrease the conduction into the mix region and therefore it can only reduce its equilibrium temperature.  However, the relevant radiative losses will depend on more detailed physics, such as the mix jet shape, x-ray line emission, the detailed temperature profile and the temporal evolution of $T_0$ and $P_0$.  Furthermore, the situation is dynamic with the high velocity jet undergoing fluid mixing. Conclusions on the full effects of magnetized mix jets will therefore require radiation-ExMHD simulations, or at the very least radiation-hydrodynamic simulations with the heat flux artificially reduced in a manner similar to Fig. \ref{fig:energy}.

\section{Discussion}
We also note several other implications of the self-generated magnetic fields.  The minimum alpha particle gyro-radius for the simulated $\mathbf{B}$ field is $60\,\mu$m. This is considerably larger than the $\simeq 5\,\mu$m size of the magnetic field regions, so the alpha deposition profile will be barely affected. However, reaction-in-flight (RIF) DT ions\cite{hayes2020plasma}, caused by large angle scattering between fast fusion products and thermal ions, may have lower energy than the fusion alpha particles and a lower gyroradius. Some of these fast ions go on to fuse, allowing diagnostic signatures of the self-generated $\mathbf{B}$ field in the high energy part of the neutron spectrum. A  comparison can be made between the angular profile of the down-scattered neutrons and that of the high energy RIF neutrons. The former result from scattering of $14\,$MeV fusion neutrons, dependent only on the assembled areal density. However, the intermediary charged DT RIF ion will also be affected by magnetic fields. 

In addition, near the ignition boundary, temperature gradients will only increase at the edge of the hot-spot. Predicted fusion burn waves\cite{michta2010effects} have core temperatures exceeding $30\,$keV and this will increase the magnetic source terms and decrease the resistive dissipation. Igniting capsules may incur significant self-generated fields of $10^4\,$T or more, which could be enough to change the alpha particle deposition profile and confine it to the hot-spot, changing the ignition burn wave dynamics. Correct modelling of igniting ICF capsules may therefore require ExMHD effects.

In summary, we have showed that carbon jets entering the fusion fuel hot-spot are especially potent for magnetization of the plasma in the $0.5\,$ns before stagnation. This is partly due to the steep temperature gradients from the increased radiative losses, and partly due to the penetration of the jet into hotter fusion plasma. We have discussed the additional thermoelectric magnetic source term that is only present for multi-species plasma. A kinetic simulation verified this source term for weakly collisional plasma with $\ln(\Lambda)\simeq 4$. Extended magneto-hydrodynamics analysis of a radiation-hydrodynamic simulation indicates that the field strength can reach $4000\,$T around a carbon jet from the fill tube perturbation. The equivalent peak magnetization exceeds 1, strong enough to decrease the electron heat conductivity by 70\%. 

In a burning plasma, heat flux into the jet is dominated by fast alpha particles and so the magnetic field will have a minimal effect on the jet during the $100\,$ps burn duration. However, the magnetization is still significant $100-500\,$ps before bang time, when fusion is insignificant and heat flux is dominated by electrons rather than alpha particles. We have seen that the fill-tube jet is the region most vulnerable to these magnetic effects. Its evolution will have a direct effect on the final hot-spot mix mass and therefore the fusion yield. A simple model of the jet as a magnetized carbon sphere identified an increased likelihood of its collapse to a dense optically thick state, even for lower mass jets. The magnetized collapse will indirectly alter hydrodynamic quantities such as the jet density, opacity, diffusivity and fluid mixing. Assessment of these effects will require full scale self-consistent radiation ExMHD simulations. 

\section*{Authors' contributions}
J.S. conducted the analysis and wrote the paper. B.H. conducted the xRAGE simulation. H.L. supervised the research.
\begin{acknowledgments}
Research presented in this article was supported by the Laboratory Directed Research and Development program of Los Alamos National Laboratory under project number 20180040DR. The authors wish to thank the scientific computing staff at Los Alamos National Laboratory. The authors acknowledge C. A. Walsh, A. C. Hayes and G. Jungman for useful discussions and D. S. Clark for provision of the implosion drive data necessary for the xRAGE simulation of shot N170601.
\end{acknowledgments}

\section*{Data Availability}
The data that support the findings of this study are available on request from the corresponding author. The data are not publicly available due export control restrictions at Los Alamos National Laboratory.


\end{document}